%% file: ieee_tnano_main.tex
\documentclass[lettersize,journal]{IEEEtran}
\usepackage[printonlyused]{acronym}
\usepackage[style=ieee,backend=bibtex,,maxbibnames=10]{biblatex}
\bibliography{biblio}

\usepackage{stfloats}
\usepackage{url}
\usepackage[font=small]{caption}
\usepackage{verbatim}
\usepackage{graphicx}
\usepackage[hidelinks]{hyperref}
\usepackage{xcolor}
\usepackage{amsmath,amsfonts}
\usepackage{algorithmic}
\usepackage{algorithm}
\usepackage{array}
\usepackage{subfigure}
\usepackage{textcomp}
\usepackage{mathtools}

\DeclarePairedDelimiter{\ceil}{\lceil}{\rceil}

\colorlet{red}{black}
\colorlet{blue}{black}

\usepackage[
  range-phrase=-,
  per-mode=symbol-or-fraction,
  range-units=single,
  list-units=single,
  detect-all,
  list-final-separator={, and },
]{siunitx}
\usepackage{silence}

\begin{document}

\title{Toward Standardized Performance Evaluation of Flow-guided Nanoscale Localization}

\author{Arnau Brosa López, Filip Lemic\IEEEauthorrefmark{1}, Gerard Calvo, Aina Pérez, Jakob Struye, Jorge Torres Gómez, Esteban Municio, Carmen Delgado, Falko Dressler, Eduard Alarcón, Jeroen Famaey, Sergi Abadal, Xavier Costa Pérez
\IEEEcompsocitemizethanks{\IEEEcompsocthanksitem\IEEEauthorrefmark{1}Corresponding author.}
\thanks{A. Brosa, A. Pérez, E. Alarcón, and S. Abadal are with Universitat Politècnica de Catalunya, Spain, email: \{name.surname\}@ac.upc.edu.}%
\thanks{F. Lemic, G. Calvo, E. Municio, C. Delgado, and X. Costa are with i2CAT Foundation, Spain, email: \{name.surname\}@i2cat.net. X. Costa is also with the NEC Laboratories Europe GmbH, Germany and ICREA, Spain.}%
\thanks{J. Torres and F. Dressler are with Technische Universität Berlin, Germany, email: \{torres-gomez,dressler\}@ccs-labs.org.}
\thanks{J. Struye and J. Famaey are with University of Antwerp - imec, Belgium, email: \{name.surname\}@uantwerpen.be.
}%
}

\markboth{IEEE DRAFT}{Brosa\MakeLowercase{\textit{et al.}}: Toward Standardized Performance Evaluation of Flow-guided In-body Localization}

\maketitle

\begin{abstract}
Nanoscale devices with \ac{THz} communication capabilities are envisioned to be deployed within human bloodstreams. 
Such devices will enable fine-grained sensing-based applications for detecting early indications (i.e., biomarkers) of various health conditions, as well as actuation-based ones such as targeted drug delivery.
Associating the locations of such events with the events themselves would provide an additional utility for precision diagnostics and treatment. 
This vision yielded a new class of in-body localization coined under the term ``flow-guided nanoscale localization''.
Such localization can be piggybacked on THz communication for detecting body regions in which biological events were localized with the traveling time reported by nanodevices flowing with the bloodstream.
From decades of research on objective benchmarking of ``traditional'' indoor localization and its eventual standardization (e.g.,~ISO/IEC18305:2016), we know that in early stages, the reported performance results were often incomplete (e.g., targeting a subset of relevant performance metrics).
Reported results in the literature carried out benchmarking experiments in different evaluation environments and scenarios and utilized inconsistent performance indicators.
To avoid such a ``lock-in'' in flow-guided localization, we propose a workflow for standardized performance evaluation of such approaches.
The workflow is implemented in the form of an open-source simulation framework that is able to jointly account for the mobility of the nanodevices, in-body THz communication with on-body anchors, and energy-related and other technological constraints (e.g., pulse-based modulation) at the nanodevice level.
Accounting for these constraints, the framework can generate raw data to streamline into different flow-guided localization solutions for generating standardized performance benchmarks. 
\vspace{-1mm}
\end{abstract}


\input{acronym_def}
\vspace{-2mm}
\input{introduction}

\input{related_works}

\input{system_overview}

\input{results}

\input{conclusion}

\vspace{-0.5mm}
\section*{Acknowledgments}

This work was supported by the European Union’s Horizon Europe (grant 101139161 — INSTINCT project); the Spanish Ministry of Economic Affairs with EU—NextGeneration EU, under the PRTR (Call UNICO I+D 5G 2021, grant TSI-063000-2021-6-Open6G); and the German Research Foundation (DFG, NaBoCom project, grant DR 639/21-2).

\vspace{-0.5mm}
\renewcommand{\bibfont}{\footnotesize}
\printbibliography

\end{document}

%% file: acronym_def.tex

\acrodef{THz}{Terahertz}
\acrodef{RF}{Radio Frequency}
\acrodef{IMU}{Inertial Measurement Unit}
\acrodef{GPS}{Global Positioning System}
\acrodef{GNN}{Graph Neural Network}
\acrodef{MAC}{Medium Access Control}
\acrodef{CSV}{Comma Separated Value}
\acrodef{ZnO}{Zinc-Oxide}
\acrodef{EVARILOS}{Evaluation of RF-based Indoor Localization Solutions for the Future Internet}
\acrodef{PerfLoc}{Performance Evaluation of Smartphone Indoor Localization Apps}
\acrodef{SRS}{Simple Random Sampling}
\acrodef{SSRS}{Stratified Simple Random Sampling}
\acrodef{CRS}{Cluster Random Sampling}
\acrodef{RGS}{Regular Grid Sampling}
\acrodef{SCS}{Spatial Coverage Sampling}
\acrodef{CPU}{Central Processing Unit}
\acrodef{MIMO}{multiple-input multiple-output}
\acrodef{SISO}{single-input single-output}
\acrodef{BVS}{BloodVoyagerS}
\acrodef{SINR}{Signal to Interference plus Noise Ratio}

%% file: introduction.tex
\section{Introduction}

Advances in nanotechnology are paving the way toward nanodevices with integrated sensing, computing, and data and energy storage capabilities~\cite{jornet2012joint}.
Among others, such devices will find applications in precision medicine~\cite{abbasi2016nano,lemic2021survey, chude2017molecular}.
A subset of such applications envisions the nanodevices being deployed in the patients' bloodstreams \cite{mosayebi2018early, moser2024}.
As such, they will have to abide to the environmental constraints limiting their physical size to the one of the red blood cells (i.e., smaller than 5~microns).
Due to constrained sizes, their sole powering option will be to scavenge environmental energy (e.g., from heartbeats or through ultrasound power transfer) utilizing nanoscale energy-harvesting entities such as \ac{ZnO} nanowires~\cite{jornet2012joint} or through chemical reactions with enzymes from the environment \cite{chen2024enzymatic}.
Due to constrained powering, such devices are expected to be passively flowing within the patients' bloodstreams. 

Recent advances in the development of novel materials, primarily graphene and its derivatives~\cite{abadal2015time}, herald nanoscale wireless communications in the $\si{\tera\hertz}$ region (i.e., $\SIrange{0.1}{10}{\tera\hertz}$)~\cite{lemic2021survey}.
In the context of the above-discussed nanodevices, wireless communication capabilities will enable their two-way communications with the outside world~\cite{dressler2015connecting, canovas2020understanding, yao2022fgor, yang2020comprehensive}.
Deployment spans from \ac{SISO}~\cite{yein2022biomedical,sangwan2022joint} to \ac{MIMO}~\cite{belaoura2021cooperative,he2014invivo} schemes for enhanced communication capabilities.
Communication-integrated nanodevices are paving the way toward sensing-based applications such as oxygen sensing in the bloodstream for detecting hypoxia (i.e., a biomarker for cancer diagnosis), and actuation-based ones such as non-invasive targeted drug delivery for cancer treatment~\cite{chen2024enzymatic}.

As recognized in the literature, communication-enabled nanodevices will also provide a primer for flow-guided localization in the bloodstreams~\cite{lemic2021survey,lemic2022toward}. 
Such localization will enable associating the location of an event detected by the nanodevices (e.g., hypoxia or a target for targeted drug delivery), providing medical benefits along the lines of non-invasiveness, early and precise diagnostics, and reduced costs~\cite{simonjan2021body,gomez2022nanosensor,lemic2022toward,vasisht2018body}. 
Flow-guided localization is in an early research phase, with only a few works targeting the problem~\cite{simonjan2021body,gomez2022nanosensor,lemic2022toward}.
The main challenges include i) a sub-centimeter range of $\si{\tera\hertz}$ in-body wireless communication at nanoscale, ii) energy-related constraints stemming from energy-harvesting as the sole powering option of the nanodevices, iii) high mobility of the nanodevices in the bloodstreams, with their speeds reaching 20~cm/s.
Flow-guided localization proposals have made encouraging progress in addressing the above challenges, yet we argue that the research and further advances on such localization are needed and yet to flourish.

Based on the above arguments and the knowledge generated through decades of research on ``traditional'' indoor localization, we posit that, at this early stage, there is a need for a framework for objective performance evaluation of flow-guided localization. 
Specifically, early research on traditional indoor localization suffered from the inability to objectively compare the performance of different approaches.
In other words, the reported performance results were often incomplete (e.g., targeting a single metric such as localization accuracy and ignoring the other important ones such as the latency of reporting location estimates), utilizing different performance indicators (e.g., mean vs. median accuracy), and utilizing different evaluation environments and scenarios.
These issues were eventually recognized in the community and addressed through large-scale and costly projects such as the EU \ac{EVARILOS}~\cite{van2015platform} and NIST \ac{PerfLoc}~\cite{moayeri2016perfloc,moayeri2018perfloc}.
Other examples are with indoor localization competitions such as the one from Microsoft at the ACM/IEEE IPSN conference~\cite{lymberopoulos2015realistic}, eventually resulting in the development of an ISO/IEC standard for objective benchmarking of indoor localization approaches~\cite{ISO/IEC18305:2016}.  

The general objective of this article is to avoid the initial ``lock-in'' in the comparability of flow-guided localization by proposing a framework for standardized performance evaluation of such localization approaches.
Specifically, our contributions include discussing the fundamentals of flow-guided nanoscale localization, providing an overview of existing approaches, and discussing the limitations of their current performance assessments.
This is followed by proposing a workflow for standardized and objective performance assessment of flow-guided localization.
In addition, an open-source network simulation framework is provided that implements the discussed workflow and provides the community with the first tool for realistic and objective assessment of flow-guided localization.
We follow by demonstrating the capabilities of the proposed simulator by evaluating the performance of a state-of-the-art flow-guided localization solution.
Finally, we provide guidelines on the optimal sampling of evaluation locations in the cardiovascular system that guarantees objective benchmarking and showcase the benefits of parallelized execution of benchmarking experiments.

%% file: related_works.tex
\vspace{-1mm}
\section{Related Works}
\label{sec:related_works}

\subsection{Flow-guided Localization Fundamentals}

The main objective of flow-guided localization is to utilize the nanodevices to localize target events. 
The work in~\cite{lemic2022toward} proposes a multi-hopping based in-body localization approach that can conceptually support flow-guided localization.
Nonetheless, the representatives of such localization are~\cite{simonjan2021body,gomez2022nanosensor}.
In these approaches, machine learning models are utilized to distinguish the body region through which each nanodevice passed during one circulation through the bloodstream.
The authors in~\cite{simonjan2021body} base this procedure on tracking the distances traversed by a nanodevice in its circulations through the bloodstream using a conceptual nanoscale \ac{IMU}.
However, this poses challenges in terms of resources available at the nanodevice level for storing and processing \ac{IMU}-generated data, and challenges related to the vortex flow of blood negatively affecting the accuracy of \ac{IMU} readings.
In~\cite{gomez2022nanosensor}, these issues are mitigated by tracking the time needed for each circulation through the bloodstream.
The captured distances or times are then envisioned to be reported to a beaconing anchor deployed near the heart, utilizing ultrasonic or short-range $\si{\tera\hertz}$-based backscattering at the nanodevice level.

Given that only a body region through which the nanodevice traversed is being detected, these localization approaches are (in contrast to~\cite{lemic2022toward}) not designed to provide point localization of the target.
This is despite the fact that point localization of the target event would be immensely beneficial for the healthcare diagnostics. 
Moreover, the region detection accuracy and reliability of localization can intuitively be enhanced with an increase in the nanodevices' number of circulations in the bloodstream. 
As a trade-off, such an increase would negatively affect the energy consumption of the localization procedure.
Therefore, in flow-guided localization, relevant performance metrics such as point and region accuracies, reliability, and energy consumption should be considered a function of the application-specific delay allowed for localizing target events.

\subsection{Performance Evaluation of $\si{\tera\hertz}$ Nanoscale Systems}

As argued in~\cite{geyer2018bloodvoyagers}, simulating the performance of a given system allows for completely controllable experimental conditions and environments. 
Combined with repeatability and cost-efficiency, these advantages make simulations a valuable tool for evaluating new algorithms, especially at early research stages.
Given that the research on flow-guided localization is still in the preliminary stage, simulating the operation of such systems can be considered a natural first step in assessing their performance.  

\begin{figure*}[ht]
\centering
\includegraphics[width=\linewidth]{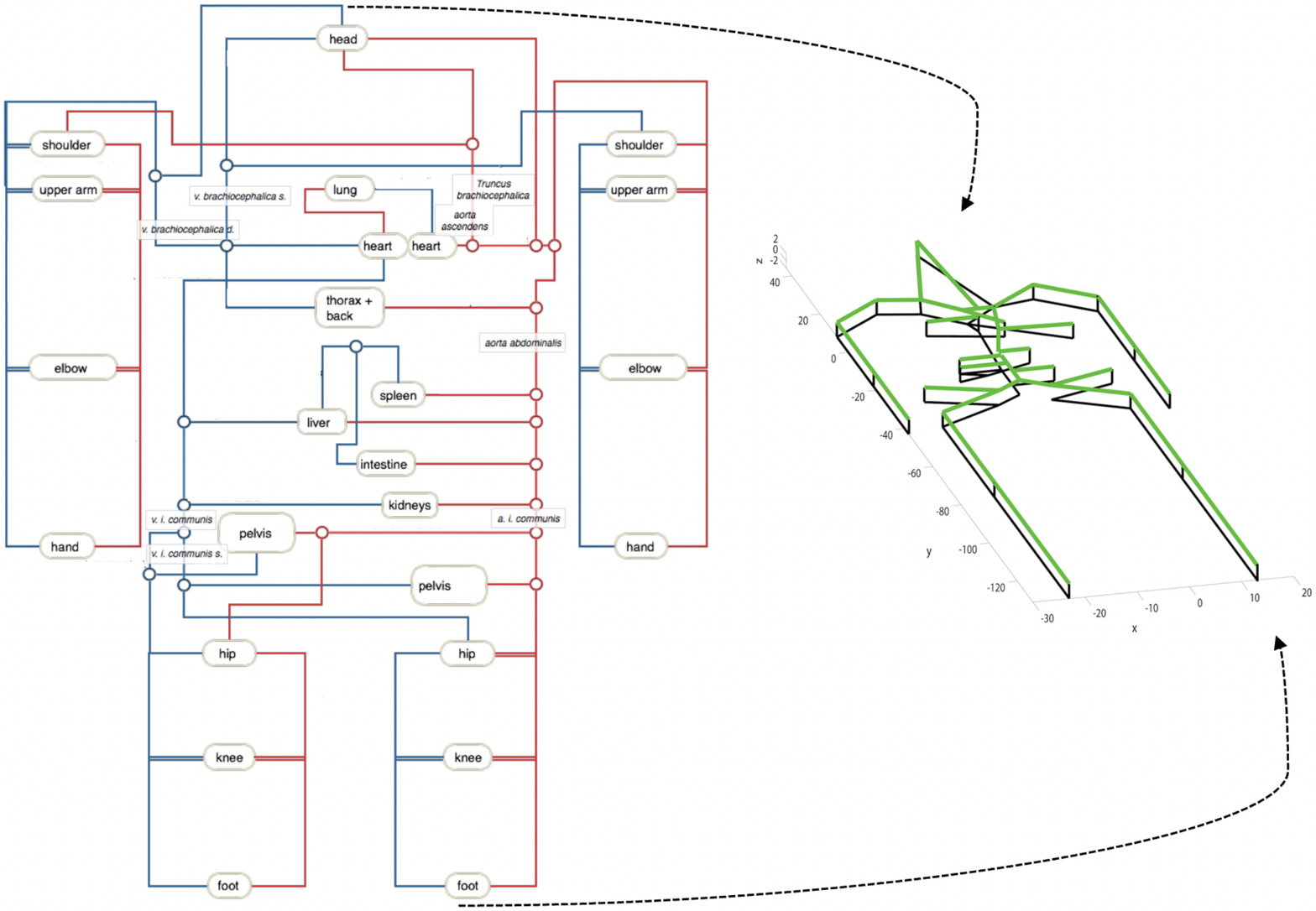}
\vspace{-3mm}
\caption{Nanodevice mobility in the \ac{BVS} (reproduced from~\cite{geyer2018bloodvoyagers})}
\label{fig:bloodvoyagers}
\vspace{-4mm}
\end{figure*}

This was only meagerly recognized in the scientific community, with \ac{BVS}~\cite{geyer2018bloodvoyagers} being the first tool that provides a simplified bloodstream model for \textit{simulating the mobility of the nanodevices}.
The simulator covers \num{94} vessels and organs, and the coordinate system's origins are placed in the heart's center.
The spatial depth of all organs is equated, with the reference thickness of $\SI{4}{\centi\meter}$ mimicking the depth of a kidney, resulting in the z-coordinates of the nanodevices being in the range from $-2$ to $\SI{2}{\centi\meter}$ (cf., Figure~\ref{fig:bloodvoyagers}). 

The simulator further assumes that the arteries and veins are set anterior and posterior, respectively. 
Transitions from the arteries to veins happen in the organs, limbs, and head.
In the heart, the blood transitions from the veins to arteries, i.e., the blood model transitions from posterior to anterior. 
The flow rate is modeled through the relationship between pressure difference and flow resistance. 
This results in the average blood speeds of \num{20}, \num{10}, and $\SIrange{2}{4}{\centi\meter\per\second}$ in aorta, arteries, and veins, respectively.
Transitions between the arteries and veins are simplified by utilizing the constant velocity of 1~cm/s.

TeraSim~\cite{hossain2018terasim} is the first simulation platform for \textit{modeling $\si{\tera\hertz}$ communication networks} which captures the
capabilities of nanodevices and peculiarities of in-air $\si{\tera\hertz}$ propagation. 
TeraSim is built as a module for ns-3 (i.e., a discrete-event network simulator), implementing physical and link layer solutions tailored to nanoscale $\si{\tera\hertz}$ communications. 
Specifically, at the physical layer, it features pulse-based communications with an omnidirectional antenna over distances shorter
than $\SI{1}{\meter}$, assuming an almost $\SI{10}{\tera\hertz}$ wide transmission window.
At the link layer, TeraSim implements two well-known protocols, i.e., ALOHA and CSMA, while a common $\si{\tera\hertz}$ channel module implements a frequency selective channel model, assuming in-air wireless communication.
We will utilize \ac{BVS} and TeraSim as \textit{the starting points} in developing the simulation framework envisioned in this work.

\subsection{Evaluation Methodologies for Flow-guided Localization}

As argued, research lessons on the performance evaluation of indoor localization systems can, to some extent, be applied to objective and standardized assessment of flow-guided localization.
The EU EVARILOS project was among the early efforts aiming at such performance assessment for RF-based indoor localization~\cite{van2015platform}.
Within the project, a performance assessment methodology was developed, which included a number of evaluation scenarios, envisioned capturing the performance of evaluated solutions along a heterogeneous set of metrics including localization accuracy, latency, and energy consumption, and assessing and mitigating the negative effects of RF interference on the performance of the evaluated solutions.
The project also yielded a web platform populated with raw data envisioned to be inputted in an indoor localization solution for its streamlined performance assessment along a number of standardized scenarios.
A similar approach was followed in the NIST PerfLoc project, however with a set of possible solutions to be evaluated extending beyond only \ac{RF} to \ac{IMU}-based, \ac{GPS}-supported, and other hybrid approaches. 
Finally, the IPSN/Microsoft Indoor Localization Competition~\cite{lymberopoulos2015realistic} was the first effort to support back-to-back evaluation of different indoor localization approaches along the same set of experimental conditions.

The above-discussed and consequent efforts yielded the following lessons: i) performance comparison of different indoor localization approaches can be carried out in an objective way by following the same evaluation methodology, i.e., utilizing the same environments, scenarios, and evaluation metrics, ii) such evaluation can be streamlined by providing a set of raw data captured along a standardized evaluation methodology, which is envisioned to be utilized as an input to an indoor localization solution under consideration, and iii) the performance of \ac{RF}-based indoor localization can be degraded by both self-interference and interference from neighboring RF-based systems operating in the same frequency band.   

In the current outlook on the performance assessment of flow-guided localization, the approaches from~\cite{simonjan2021body,gomez2022nanosensor} are evaluated in a rather simplified way accounting solely for the mobility of the nanodevices as modeled by the \ac{BVS}. 
As such, these assessments ignore many potential effects of wireless communication (e.g., \ac{RF} interference), as well as energy-related constraints stemming from energy-harvesting and, consequently, the intermittent nanodevice operation~\cite{jornet2012joint}.
It is also worth mentioning that authors in~\cite{lemic2022toward} carried out a limited performance evaluation assessing the number of nanodevices needed for localizing a nanodevice at any location in the body in a multi-hop fashion.
The derived assessments can, therefore, at this point only serve as a rough indication due to their low levels of realism and subjective evaluation methodologies. 
In this work, we enhance the realism of such assessments by jointly accounting for the mobility of the nanodevices, in-body nanoscale $\si{\tera\hertz}$ communication between the nanodevices and the outside world, and energy-related and other technological constraints (e.g., pulse-based modulation) of the nanodevices.

\begin{figure*}[!t]
	\centering
	\includegraphics[width=\linewidth]{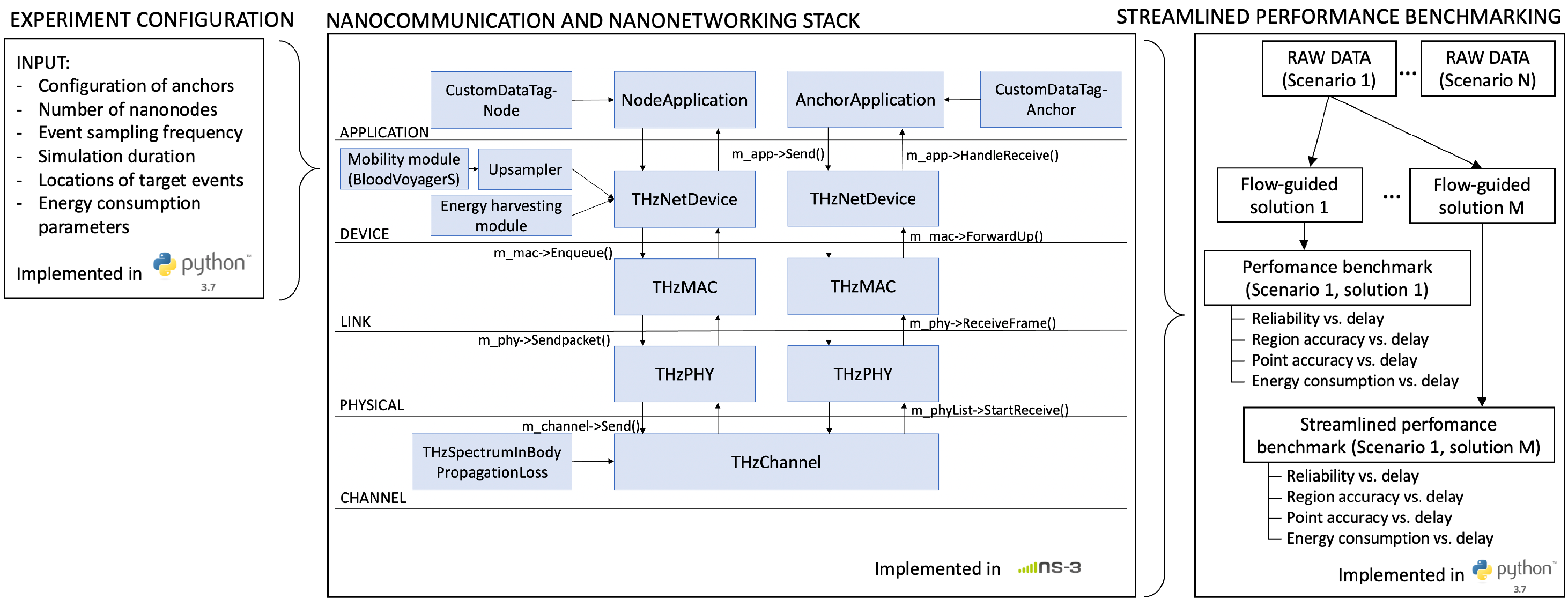}
	\vspace{-1mm}
	\caption{Overview of the framework for standardized performance evaluation of flow-guided localization}
	\label{fig:simulator}
	\vspace{-3mm}
\end{figure*}

%% file: system_overview.tex
\section{Framework for Standardized Performance Evaluation of Flow-guided Localization}

\subsection{Evaluation Workflow}

As mentioned, enabling flow-guided localization in the bloodstream requires at least a single anchor mounted on the patient's body.
Flow-guided localization approaches in~\cite{gomez2022nanosensor,simonjan2021body} can be enabled with a single anchor strategically positioned in the proximity of the heart. 
This is because the heart is the only location through which each nanodevice is guaranteed to pass in each circulation through the bloodstream.
Additional anchors can be introduced into the system by specifying their coordinates in their configuration file of the simulator, as indicated in Figure~\ref{fig:simulator}. 
The on-body anchors are expected to feature batteries or similar powering sources, hence they are assumed to be continuously operational.
Their main roles are transmitting beacon packets and receiving the nanodevices' backscattered responses. 

\textcolor{red}{
The nanodevices are assumed to feature capacitors for energy storage and ZnO nanowires as the energy-harvesting entities~\cite{wang2008energy}, where the energy is harvested in nanowires' compress-and-release cycles.
The harvested energy can be specified with the duration of the harvesting cycle $t_\mathrm{cycle}$ and the harvested charge per cycle $\Delta Q$.
The capacitor charging is modeled as an exponential process accounting for the energy-harvesting rate and interval (e.g., $\SI{6}{\pico\joule}$ per second and per 20~ms for harvesting from heartbeats and ultrasound-based power transfer, respectively~\cite{jornet2012joint}) and capacitor's storage capacity. 
Specifically, the model accounts for the total capacitance of the nanonode, denoted as $C_\mathrm{cap}$, and evaluated as \mbox{$C_\mathrm{cap} = 2 E_\mathrm{max} / V_g^2$}, i.e., $C_\mathrm{cap}$ depends on the energy storage capacity $E_\mathrm{max}$ and the generator voltage $V_g$.  
In the modeling, it is required to know in which harvesting cycle the nanonode is, as denoted by $n_\mathrm{cycle}$, and given its current energy level $E_{n_\mathrm{cycle}}$, which can be derived from~\cite{jornet2012joint} as follows:}

\begin{equation}
\label{eq1}
n_\mathrm{cycle} = \ceil*{\frac{- V_g C_\mathrm{cap}}{\Delta Q} ln\left(1 - \sqrt{\frac{2 E_{n_\mathrm{cycle}}}{C_\mathrm{cap} V_g^2}}\right)}.
\end{equation}

\textcolor{red}{The energy in the next energy cycle $n_\mathrm{cycle} + 1$ is then: }

\begin{equation}
\label{eq2}
E_{n_{\mathrm{cycle}+1}} = \frac{C_\mathrm{cap} V_g^2}{2} \left(1- e^{-\frac{\Delta Q (n_\mathrm{cycle} + 1)}{V_g C_\mathrm{cap}}}\right)^2.
\end{equation}

\begin{figure}[!t]
	\centering
	\includegraphics[width=\linewidth]{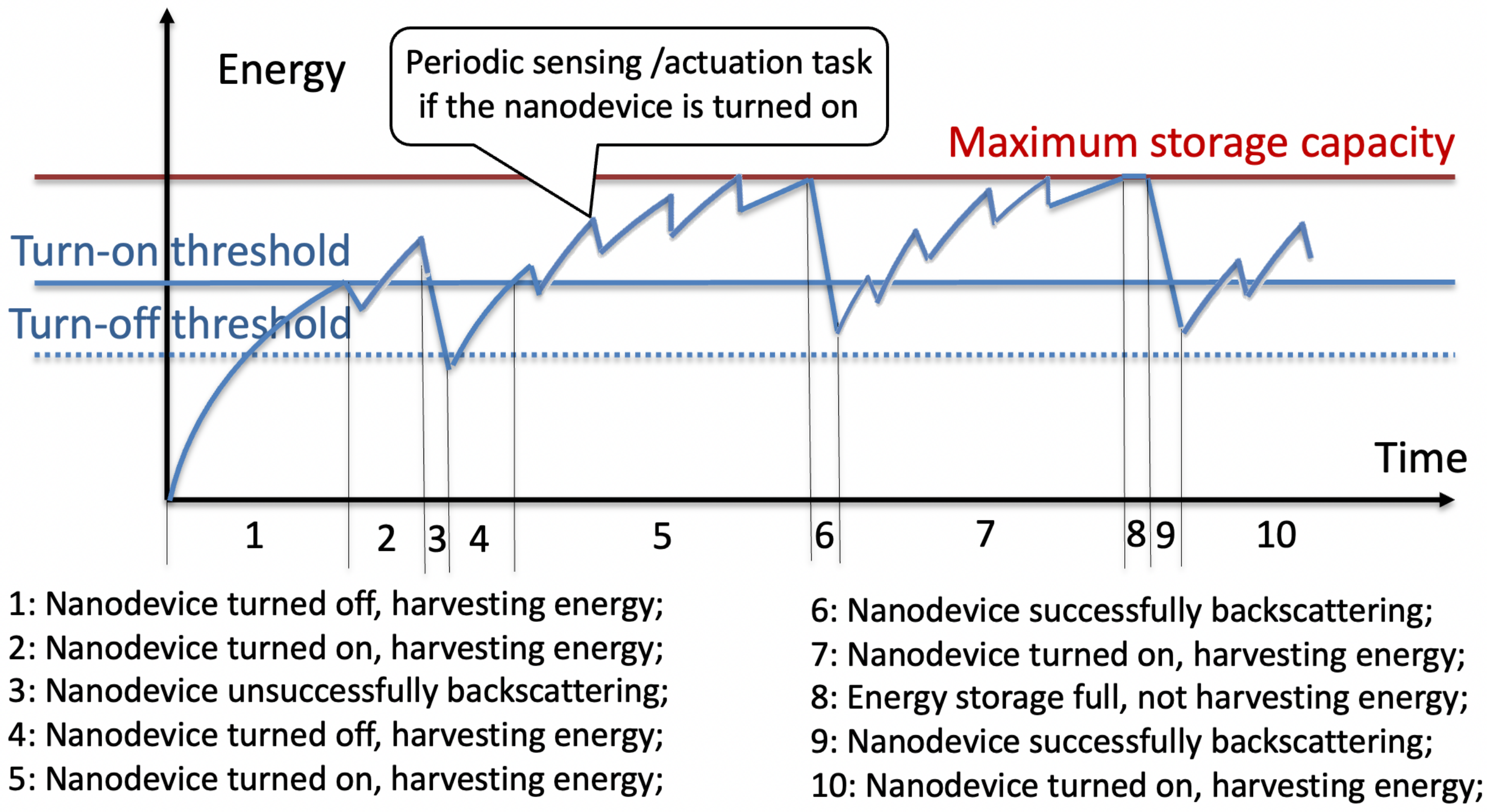}
	\vspace{-1mm}
	\caption{Lifecycle of an energy-harvesting nanodevice}
	\label{fig:energy_harvesting}
	\vspace{-3mm}
\end{figure}

The nanodevices are assumed to feature intermittent behavior due to harvesting and storage constraints.
In other words, once its energy falls below the \emph{Turn OFF} threshold, the nanodevice turns off, followed by a turn on when its energy increases above the \emph{Turn ON} threshold, \textcolor{red}{as shown in Figure~\ref{fig:energy_harvesting}}.
If the nanodevices are turned on, they are assumed to periodically carry out a sensing or actuation task with a given frequency.
Each task execution is expected to consume a certain constant amount of energy; hence, the more frequent the task, the more energy each nanodevice will consume.

The target event location(s) is (are) envisioned to be hard-coded by the experimenter, abiding by the scenario's constraints. 
Specifically, this location has to be in or near the bloodstream to eventually be detected by the nanodevices.
The event is assumed to be detected by a nanodevice if i) the Euclidean distance between its location and the one of the nanodevice at the time of the task execution smaller than the predefined threshold (n.b., configured to $\SI{1}{\centi\meter}$ in the reported experiments), and ii) the nanodevice is turned on.
\textcolor{red}{The reasoning behind such modeling is in the fact that diagnostically relevant events are assumed to be static, and releasing into the bloodstream certain biomarkers for event detection.
The released biomarkers are assumed to feature a lifetime substantially shorter than the deployment of the nanodevices in the bloodstream, effectively creating a spherical region around the event with the radius equalling the predefined event detection threshold. 
Within the region, the nanodevice is able to sense the biomarkers for event detection.}

Communication with the anchor is based on passive reception of a beacon, followed by active (i.e., energy-consuming) transmission of a response packet from the nanodevice, as assumed in the representative work from the literature~\cite{gomez2022nanosensor}. 
The anchor is beaconing with the constant beaconing frequency and transmit power. 
In each beacon packet, the anchor advertises its \ac{MAC} address.
In the backscattered packets, the nanodevices report their \ac{MAC} addresses, the time elapsed since their last passage through the heart, and an event bit.
The time elapsed since the last passage through the heart and the event bit represent the raw data that can be fed into a flow-guided localization approach for localizing a target event.  
Each time a nanodevice passes through the heart, the time elapsed since the last passage is re-initialized to zero to not compound multiple circulations.
The event bit is assumed to be a logical ``1'' in case of successful detection of a target event and ``0'' otherwise.
Similarly, the event bit is reinitialized to ``0'' in each passage through the heart.

\subsection{Framework Design and Implementation}

The simulation framework for standardized performance evaluation of flow-guided localization is depicted in Figure~\ref{fig:simulator}.
The input to the framework is a set of parameters defining an evaluation scenario.
The inputs are envisioned to be passed to the ns-3-based simulator for the generation of raw data to be used for streamlined evaluation of a given flow-guided localization solution for the assumed scenario, resulting in a performance benchmark, as indicated on the right hand side in Figure~\ref{fig:simulator}.
Each streamlined performance benchmark consists of a set of relevant performance metrics, in turn allowing for an objective back-to-back comparison of different approaches in a consistent environment along the same set of scenarios and performance metrics.
\textcolor{red}{Specifically, as the relevant performance metrics we consider the point and region localization accuracy, reliability of providing location estimates, and energy consumption of the nanodevices.
Region accuracy is calculated as the percentage of correctly estimated regions:}

{\small
\begin{equation}
\mathrm{Region}\_\mathrm{acc.} \ [\%] = \frac{\textcolor{blue}{N_{\text{correct}}}}{\textcolor{blue}{N_{\text{total}}}},
\end{equation}}
where \textcolor{blue}{$N_{\text{correct}}$} represents the number of correct region estimates, and \textcolor{blue}{$N_{\text{total}}$} the total number of evaluation points.

\textcolor{red}{
The point accuracy for each evaluation point is calculated as the Euclidean distance between the true location $(x_T,y_T,z_T)$ of the event hard-coded by the experimenter and the estimated location or region centroid $(x_E,y_E,z_E)$ reported by the flow-guided localization approach:}

{\small
\begin{equation}
\mathrm{Point}\_\mathrm{acc.} \ [\si{\centi\meter}] = \sqrt{(x_{E}-x_{T})^2 + (y_{E}-y_{T})^2 + (z_{E}-z_{T})^2}.
\end{equation}}

The simulator's architecture follows a well-established ns-3 layered model, as depicted in Figure~\ref{fig:simulator}.
The \emph{AnchorApplication} module implements continuous beaconing with a predefined period (n.b., with $\SI{100}{\milli\second}$ being a default value). 
Each beacon packet is forwarded to the \emph{THzNetDevice} module toward the communication stack implemented within the TeraSim simulator.
The link and physical layers implement the ALOHA protocol and TS-OOK modulation, respectively. 

The $\si{\tera\hertz}$ channel is modeled by calculating the receive power for each communicating pair of devices and scheduling the invocation of the \emph{ReceivePacket()} method accounting for the corresponding propagation time. 
The channel model entails in-body path-loss and Doppler terms~\cite{simonjan2021body}.
The path-loss is calculated using the attenuation and thickness parameters of the vessel, tissue, and skin. 
The Doppler term is accounted for by evaluating the change in relative positions between the nanodevices and anchors with time.
The \emph{ReceivePacket()} method checks for potential collisions by calculating the \ac{SINR} and discarding the packet if the \ac{SINR} is below the predefined threshold for reception. 
Alternatively, the packet is passed through all the way to the application layer of the nanodevice. 
At the nanodevice level, the beacon's receive power is used to set up the packet's transmission power to be backscattered.
This is followed by backscattering the response packet from the nanodevice toward the anchor, utilizing the same procedure as for transmitting the beacon. 

The anchors are assumed to be static entities and feature sufficient energy for continuous operation.
The nanodevices are assumed to be mobile energy-harvesting entities within the bloodstream.
We have integrated \ac{BVS} to model their mobility in our simulator, as visible in Figure~\ref{fig:simulator}.
Invoking a \ac{BVS} execution results in generating a \ac{CSV} file that specifies the locations of the nanodevices in the bloodstream within a simulation time frame, sampled at $\SI{1}{\hertz}$.
Since ns-3 is an event-driven simulator, at each \ac{BVS}-originating location of a nanodevice, the nanodevice is assumed to carry out a sensing/actuation task.  
Given that for certain applications carrying out such tasks could be required more frequently, we provide an upsampler for \ac{BVS}-originating locations sampled at $\SI{1}{\hertz}$.
\textcolor{red}{Each vessel in \ac{BVS} is modeled as a straight line, while the blood speed within a vessel is constant. 
The upsampling by a factor $N$ from the originally $\SI{1}{\hertz}$ sampled locations is then based on introducing additional $N-1$ nanodevice locations at equidistant times along the vessel segments.}

To each of the upsampled locations, a small random component drawn from a zero-mean Gaussian distribution \( \mathcal{E} = (x_\mathcal{E} , y_\mathcal{E} , z_\mathcal{E}) \sim N (0, \sigma^2) \) is introduced, \textcolor{blue}{which serves as our way of modeling several natural peculiarities of blood flow, including its vortex and laminar nature, as well as minor changes in the diameter of veins and arteries~\cite{kuran2020survey,jamali2019channel,wicke2018modeling}. This way of modeling the vortex and laminar nature of the blood flow with flow velocities varying across the vessel cross-sections allows us to achieve a balance between simulation accuracy and capturing the performance across the entire cardiovascular system effectively within reasonable execution times.}
The newly introduced set of locations $p_i=(x_i,y_i,z_i)$ can be obtained by defining a vector $\nu$ between the two originally sampled locations $p_0=(x_0,y_0,z_0)$ and $p_N=(x_N,y_N,z_N)$:
\begin{equation}
\nu = p_N - p_0 = (x_N - x_0, y_N - y_0, z_N - z_0).
\end{equation}
The newly introduced locations \textcolor{blue}{$p_i$} then follow as:
\begin{equation}
p_i = (x_i,y_i,z_i) = p_0 + \frac{i}{N}\frac{\nu}{||\nu||} + \mathcal{E}_i, \ \forall i \in \{1,N-1\},
\end{equation}
with $||\nu||$ being the Euclidean distance between $p_N$ and $p_0$.

\begin{figure*}[!t]
	\centering
	\includegraphics[width=0.87\linewidth]{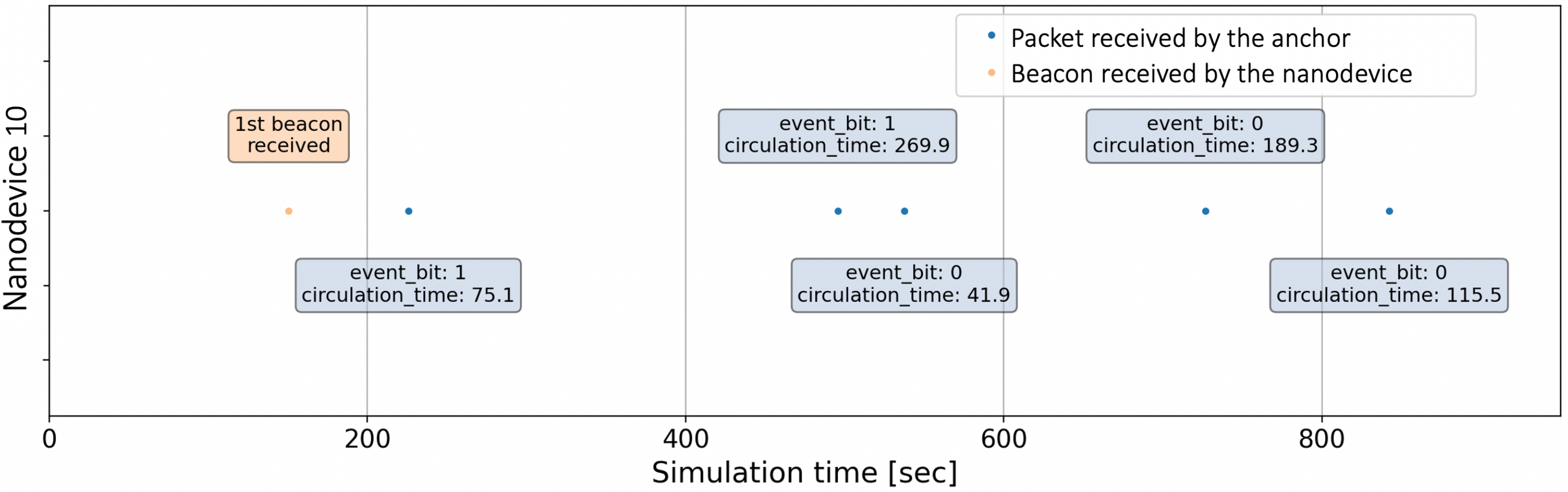}
	\caption{An example raw data output}
	\label{fig:raw_data}
	\vspace{-3mm}
\end{figure*}

%% file: results.tex
\section{Evaluation Results}

In this section, we first provide and discuss an example snapshot of the simulator-generated raw data, as well as of the performance metrics generated by the streamlined utilization of the raw data. 
We follow by assessing the performance's sensitivity to the target events' locations.
By doing so, we aim to establish the optimal method for sampling the target events in the bloodstream and the number of such samples required.
We require that the performance assessment obtained through such sampling be a reliable representation of the average performance of the considered solution within the entire bloodstream or some of its regions.
Addressing this issue is also considered a part of our future efforts.
Finally, we demonstrate the advantages of parallelizing the utilization of the simulator in terms of execution time compared to sequential execution of the same experiment. 

\subsection{A Snapshot of Framework-generated Outputs}

\begin{figure}[!t]
	\centering
	\includegraphics[width=0.98\linewidth]{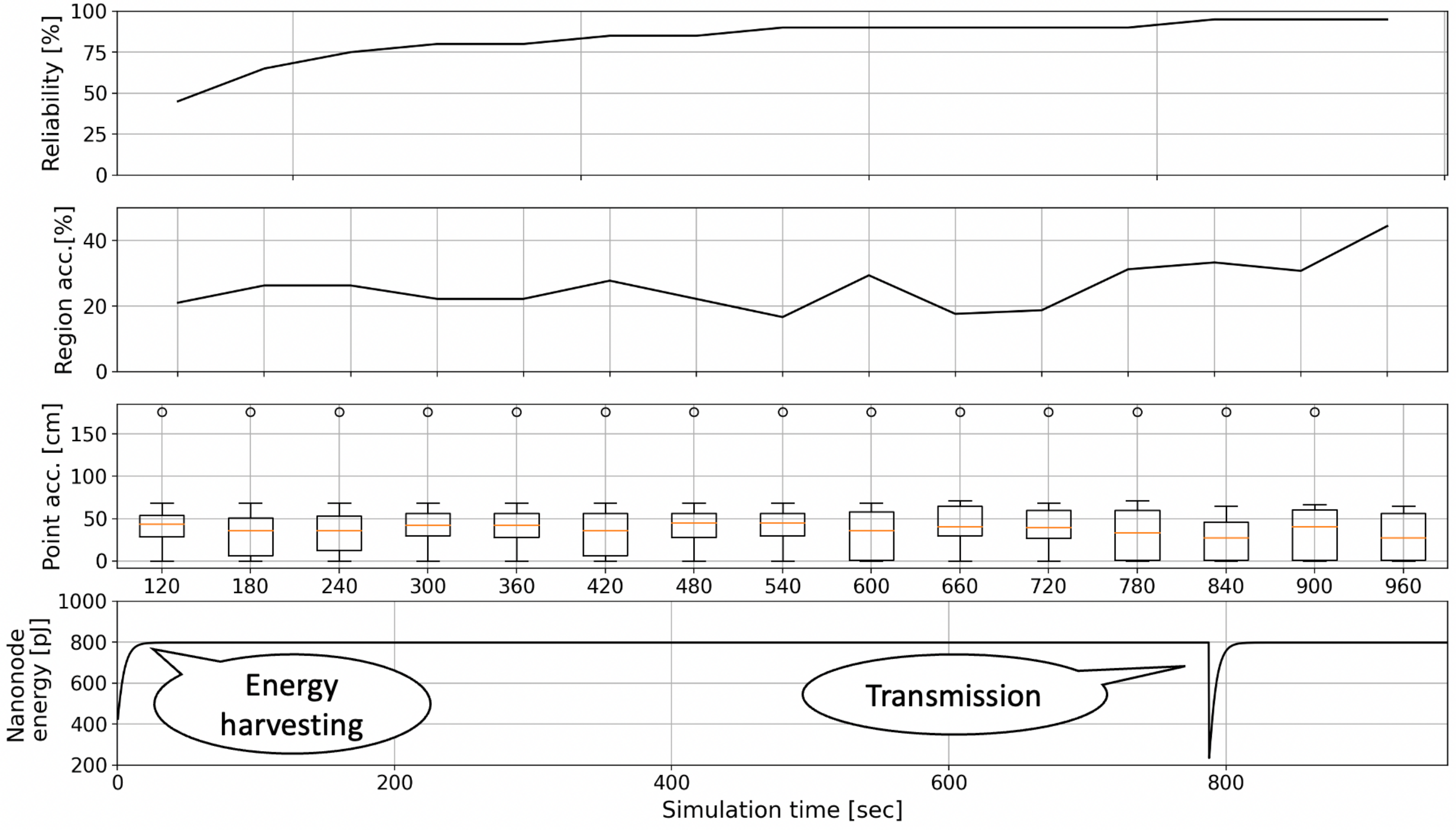}
	\vspace{-1mm}
	\caption{An example streamlined performance benchmark}
	\label{fig:example_results}
	\vspace{-3mm}
\end{figure}

A snapshot of the framework-generated outputs is depicted in Figures~\ref{fig:raw_data} and~\ref{fig:example_results}.
In the generation of the outputs, we have utilized a single anchor positioned in the center of the heart, \num{64} nanodevices sampling for target events at \num{3} samples per second, ultrasound-based energy-harvesting at the nanodevice level~\cite{jornet2012joint}, the overall simulation duration of $\SI{1000}{\second}$, and the Euclidean distance for detecting a target event of $\SI{1}{\centi\meter}$.
\textcolor{blue}{
The baseline simulation parameters used in our study are based on~\cite{jornet2012joint} and summarized in Table~\ref{tab:baseline_parameters}.}

\begin{table}[!t]
\centering
\caption{Baseline Simulation Parameters}
\begin{tabular}{l l }
\hline
\textbf{Parameter} & \textbf{Value} \\ \hline
Anchor beaconing interval & 100 ms \\ 
Nanodevice sampling rate & 3 samples per second \\
Simulation duration & 1000 s \\ 
Event detection distance & 1 cm \\ 
Blood flow speed (aorta) & 20 cm/s \\ 
Blood flow speed (arteries) & 10 cm/s \\ 
Blood flow speed (veins) & 2 - 4 cm/s \\ 
Transition speed (organs/limbs/head) & 1 cm/s \\
Generator voltage \( V_g \) [V] & 0.42 \\ 
Energy consumed in pulse reception [pJ] & 0.0 \\
Energy consumed in pulse transmission [pJ] & 1.0 \\
Maximum energy storage capacity [pJ] & 800 \\ 
Turn ON/OFF thresholds [pJ] & 10/0 \\
Harvesting cycle duration [ms] & 20 \\
Harvested charge per cycle [pC] & 6 \\ 
Transmit power \( P_{TX} \) [dBm] & -20 \\ 
Operational bandwidth [GHz] & 10 \\ 
Receiver sensitivity [dBm] & -110 \\ 
Operational frequency [THz] & 1 \\ \hline
\end{tabular}
\label{tab:baseline_parameters}
\vspace{-5mm}
\end{table}

Figure~\ref{fig:raw_data} depicts the raw data generated by an example nanodevice during one simulation runtime. 
The raw data consists of the \emph{circulation\_time} parameter indicating the time passed since the last reception of a beacon from the anchor and the \emph{event\_bit} suggesting if the target event was detected since the last beacon reception.
The main takeaway from Figure~\ref{fig:raw_data} is that, for some raw data instances, the \emph{circulation\_time} is larger than $\SI{90}{\second}$, which is the maximum circulation time that might occur in a single loop through the bloodstream.
This implies that in some circulations the raw data is not reported to the anchor and, when the data is eventually reported, it contains the compound of multiple such circulations. 
Such behavior is a result of one of the following: i) intermittent operation of a nanodevice due to energy-harvesting, resulting in the nanodevice sometimes not featuring sufficient energy for sensing or transmission, and ii) self-interference from the other nanodevices and anchors, resulting in reception and transmission errors.
In addition, random paths of the nanodevices in the vicinity of the target event (i.e., in an organ, limb, or head) can result in the nanodevices missing the event due to its Euclidean distance from the event never being smaller than the threshold of $\SI{1}{\centi\meter}$, even though they went through the loop that contained the event. 
This implies that the \emph{event\_bit} parameter might in some cases be erroneous.

Figure~\ref{fig:example_results} depicts a set of performance metrics generated in a streamlined fashion using the framework.
In generating the results, we have utilized a modified approach from~\cite{gomez2022nanosensor} and 20 randomly sampled evaluation points (i.e., target events) in the bloodstream.
The modification in the approach pertains to random selection of the left or right regions given that the approach assuming a single anchor is by-design unable to distinguish between such regions for certain parts of the body (e.g., limbs).
\textcolor{red}{As visible from the figure, localization reliability increases as a function of simulation time, where the simulation time represents the duration of the administration of the nanodevices in the bloodstream.}
As an example, the localization reliability is increased from less than $\SI{50}{\percent}$ to more than $\SI{90}{\percent}$ if the simulation time is increased from \num{2} to $\SI{15}{\minute}$.
\textcolor{red}{The reason for that can be found in the fact that the prolonged duration of the nanodevices' administration in the bloodstream results in more raw data being reported to the anchor, in turn increasing the reliability of producing location estimates.} 

Our results also reveal that certain assumptions made in earlier works on flow-guided localization ignore several phenomena that are expected to occur in practice, pertaining to unreliable \ac{THz}-based communication between in-body nanodevices and on-body anchors, and intermittent operation of the nanodevices due to energy-harvesting. 
When these are accounted for as done when utilizing the proposed framework, our results reveal relatively poor performance of the evaluated flow-guided localization solution in the considered scenario. 
Specifically, the region detection accuracy is at most $\SI{40}{\percent}$ and features only a small increase with the simulation time.

Given that the approach from~\cite{gomez2022nanosensor} cannot report point estimates but solely the estimated regions, in calculating the point accuracy, we have utilized the centroid of a region as its point estimate.
\textcolor{red}{Besides, given that each region in BloodVoyagerS is modeled as a straight line with constant blood speed, its centroid is calculated as the arithmetic mean of the endpoints of the line.}
This procedure is well-established in the domain of benchmarking of proximity-based indoor localization solutions~\cite{van2015platform}.
In Figure~\ref{fig:example_results}, the depicted point accuracy can be considered irrelevant, given the low region detection accuracy.
In other words, the point accuracy should be derived only for the correctly detected regions in order to express the fine-grained ability of localizing target events.
We nonetheless depict the point accuracy even for the case of incorrectly detected regions to draw readers' attention to this issue.
The point accuracy is depicted in a regular box-plot fashion, where each box-plot depicts the distribution of localization errors for the \num{20} considered target events and a given simulation time. 
Finally, the time-dependent energy level of an example nanodevice depicted in Figure~\ref{fig:example_results} indicates the energy consumption of different tasks at the nanodevice level.
Such indications are necessary for energy-aware optimizations of the task scheduling to maximize the operational time of the intermittently-operating nanodevices in a similar way as in~\cite{jornet2012joint}.

\subsection{Sampling of Target Event Locations}

The proposed simulator is able to produce raw data that can be streamlined into flow-guided localization approaches. 
Simulation parameters are ``hard-coded'' by the user, which includes the hard-coding of the true coordinates of the target events. 
The aim of this section is to examine the extent to which the performance metrics fluctuate as a function of the true locations of the events. 
This necessity comes from observed inconsistencies in data reliability and variability across different body regions. 
For instance, events closer to the heart generally provide more consistent information than ones occurring further away, such as in the arms or legs. 

To guarantee that the reported performance metrics are representative of the entire cardiovascular system or some of its regions of interest, we need a method for selecting the true locations of target events. 
For this purpose, we assessed a number of strategies for sampling target event locations from the cardiovascular system. 
The samples generated by such strategies should be sufficiently large to provide a representative assessment in terms of region and point accuracy. 
Data generation is a time consuming process, and running a simulation on a predetermined target event is both computationally and time consuming. 
As a result, the methods should simultaneously strive to minimize the number of target event locations for striking a balance between resource efficiency and objective benchmarking.

We consider spatial sampling methods from the family of random sampling due to the fact that they ensure that the resulting sample is representative of a population~\cite{wang2012review}.
The sampling methods considered in this work are i) \ac{SRS}, i.e., randomly drawing points from the population with equal probability, ii) \ac{SSRS}, i.e., dividing the population into discrete non-overlapping strata and random from each stratum, iii) \ac{CRS}, i.e., dividing the population into clusters, followed by randomly selecting some of the clusters as a sample, iv) \ac{RGS}, i.e., sampling at regular distances following a grid pattern spanning the entire area of interest, and v) \ac{SCS}, i.e., dividing the area or volume into smaller sections, followed by taking representative samples from each.

\begin{figure}[!t]
	\centering
	\includegraphics[width=\linewidth]{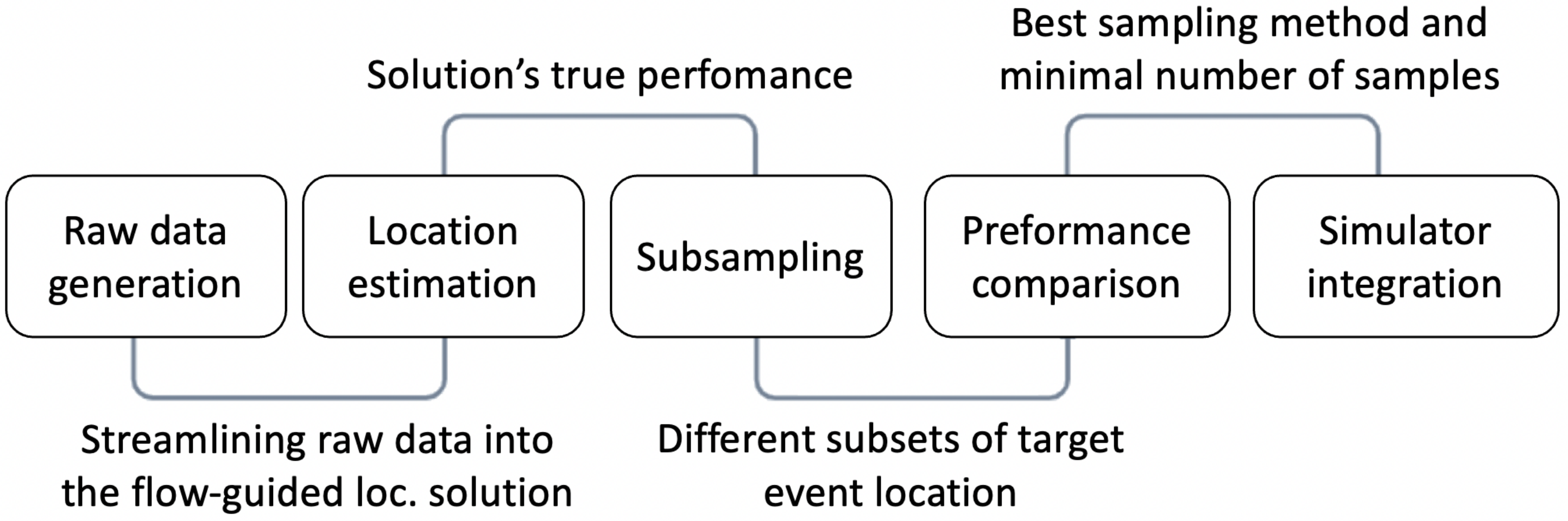}
	\vspace{-1mm}
	\caption{Methodology for deriving the optimal target event location sampling strategy and the minimal number of samples}
	\label{fig:sampling}
	\vspace{-3mm}
\end{figure}  

Our methodology for deriving the optimal sampling strategy and the minimal number of samples is depicted in Figure~\ref{fig:sampling}. 
First, a performance benchmark is generated using a densely sampled set of event locations from the cardiovascular system.
Second, the above-outlined sampling strategies are utilized to sample the original dense set of locations.
The subsets yielded by different strategies are then utilized to obtain the performance of the considered flow-guided localization solution and analyze its sensitivity to the selection of event locations. 
Lastly, the optimal sampling strategy and minimal number of samples are derived and integrated into the proposed simulation framework.

\begin{figure}[!t]
\centering
\subfigure[Region accuracy]{
\includegraphics[width=0.95\linewidth]{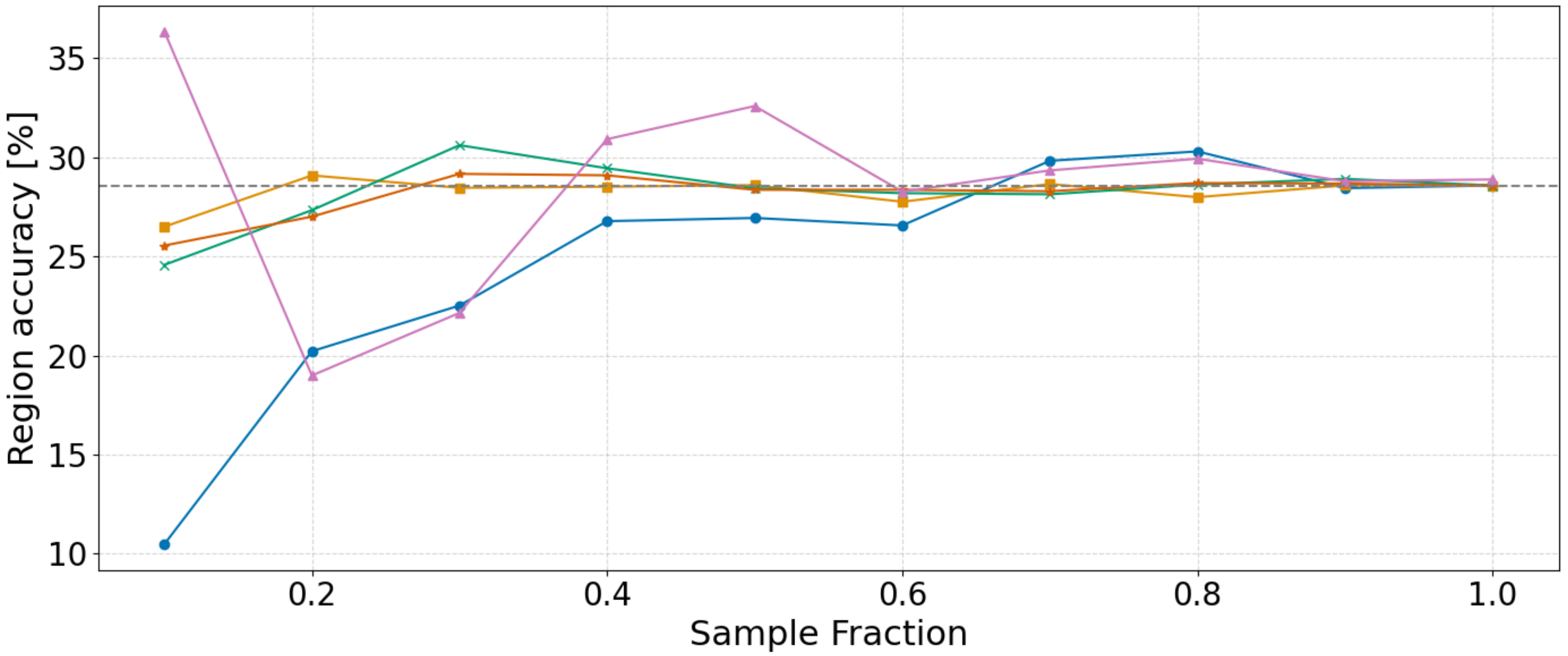}}
\subfigure[Point accuracy]{
\includegraphics[width=0.95\linewidth]{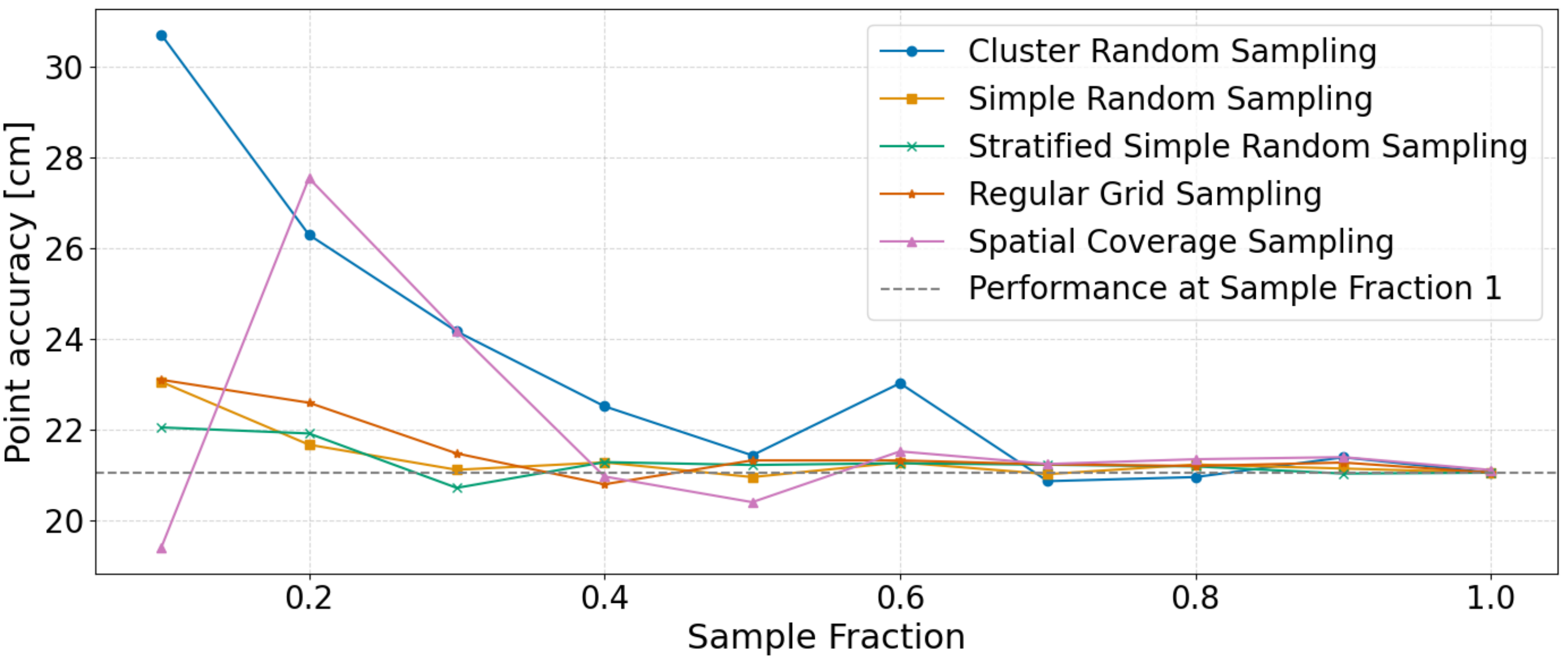}}
\vspace{-3mm}
\caption{Performance convergence of different sampling strategies}
\label{fig:sampling_convergence}
\end{figure}

The performance of the considered solution averaged over \num{10} runs of an example experiment yielded a region accuracy of around $\SI{28}{\percent}$ and point accuracy of $\SI{21.1}{\centi\meter}$, as depicted in Figure~\ref{fig:sampling_convergence}. 
The simulator assumes three different region types, i.e., \textit{Region type = 0} includes parts of the cardiovascular system with blood speeds of $\SI{20}{\centi\meter\per\second}$ (nb., aorta) and $\SI{10}{\centi\meter\per\second}$ (nb., arteries), \textit{Region type = 1} with speeds of $\SIrange{2}{4}{\centi\meter\per\second}$ (nb., veins), and \textit{Region type = 2} representing transitions between arteries and veins in organs, limbs and head, simplified with a constant velocity of $\SI{1}{\centi\meter\per\second}$.
The average performance achieved varies across different region types, with type 2 featuring the lowest accuracy of $\SI{13.2}{\percent}$ and type 1 the highest of $\SI{28.9}{\percent}$. 
The localization error is the lowest for type 1 and equals $\SI{16.8}{\centi\meter}$ and highest for type 0 with $\SI{23}{\centi\meter}$.
Differences in performance are primarily the results of different blood speeds in different regions.

\textcolor{blue}{In Figure~\ref{fig:sampling_convergence}, we also analyze the convergence of performance metrics as a function of the evaluation set size for different sampling strategies.} 
As visible, the worst-performing strategies are \ac{CRS} and \ac{SCS}. Thus, they are excluded from further consideration. 
The remaining three strategies converge toward the performance observed when utilizing an entire set of target locations at roughly $\SI{50}{\percent}$ of the utilized locations.
\textcolor{blue}{The initial fluctuations in accuracy are due to the small size of the evaluation set at the beginning of the experiment. As the number of evaluation points increases, the accuracy metric stabilizes and becomes more representative of the entire cardiovascular system. This shows the importance of selecting a sufficiently large evaluation set for consistent accuracy results.} 

\begin{figure*}[!t]
\centering
\subfigure[Dense]{
\includegraphics[width=0.238\linewidth]{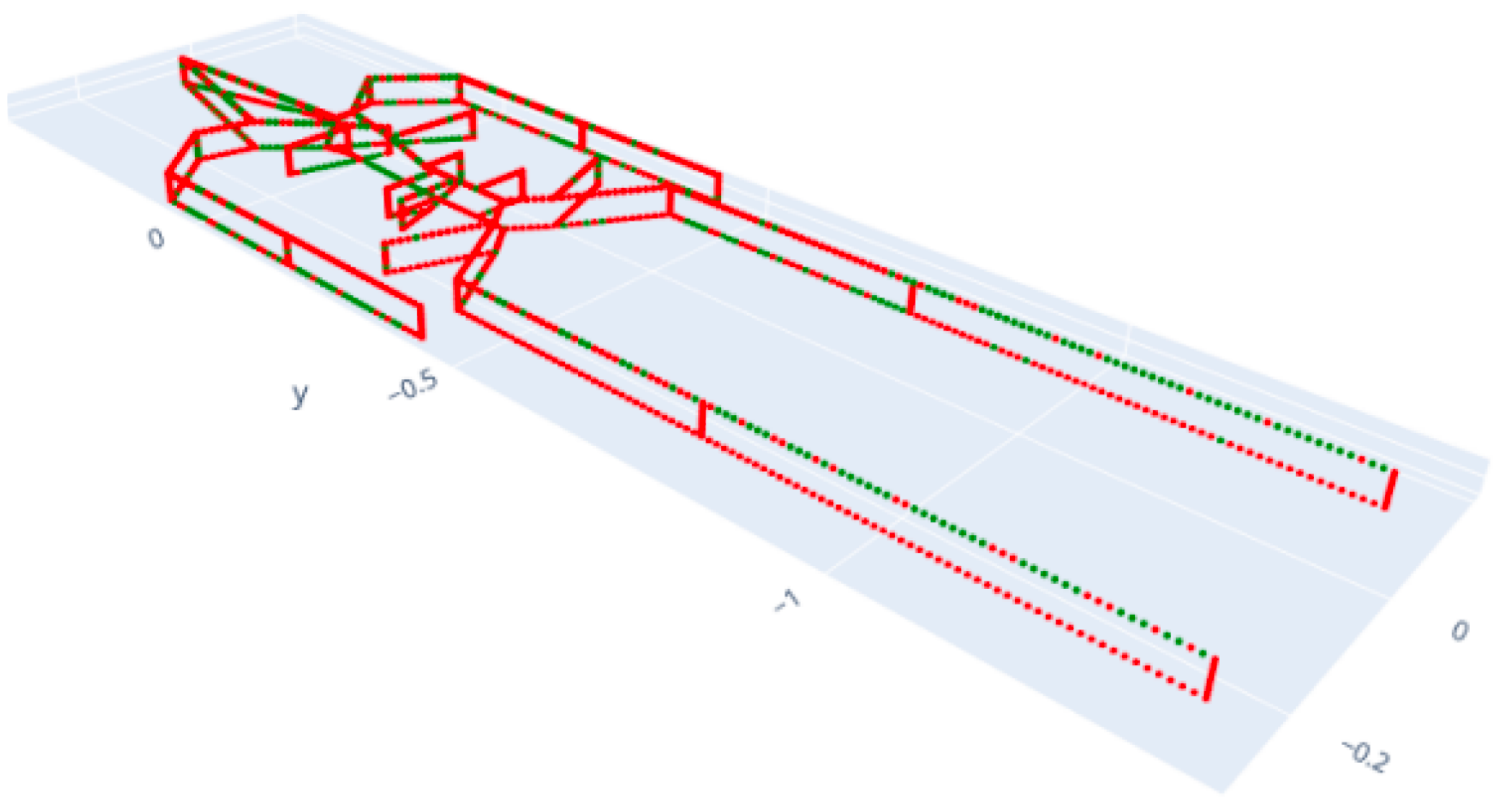}}
\subfigure[RSS]{
\includegraphics[width=0.238\linewidth]{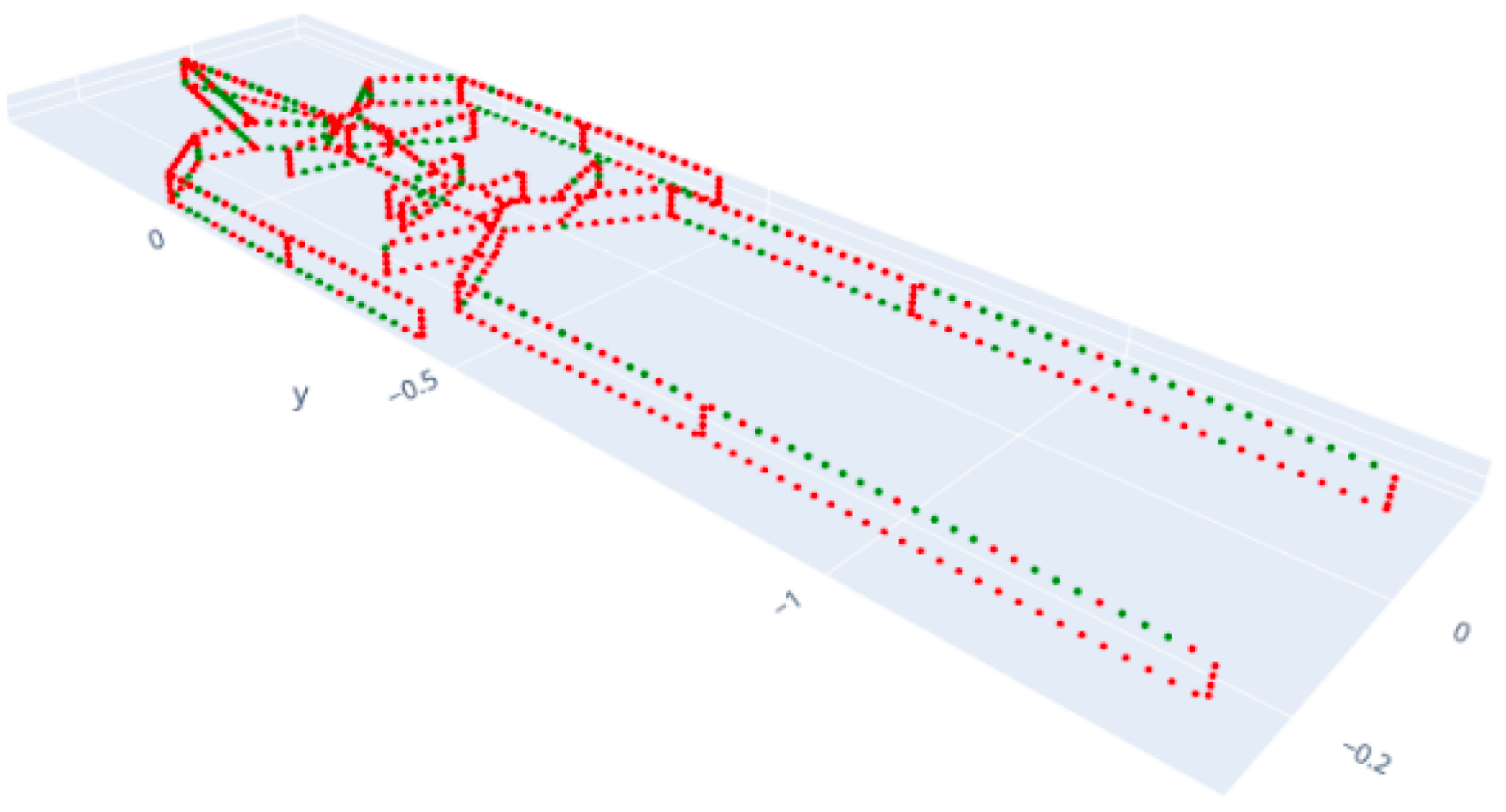}}
\subfigure[SSRS]{
\includegraphics[width=0.238\linewidth]{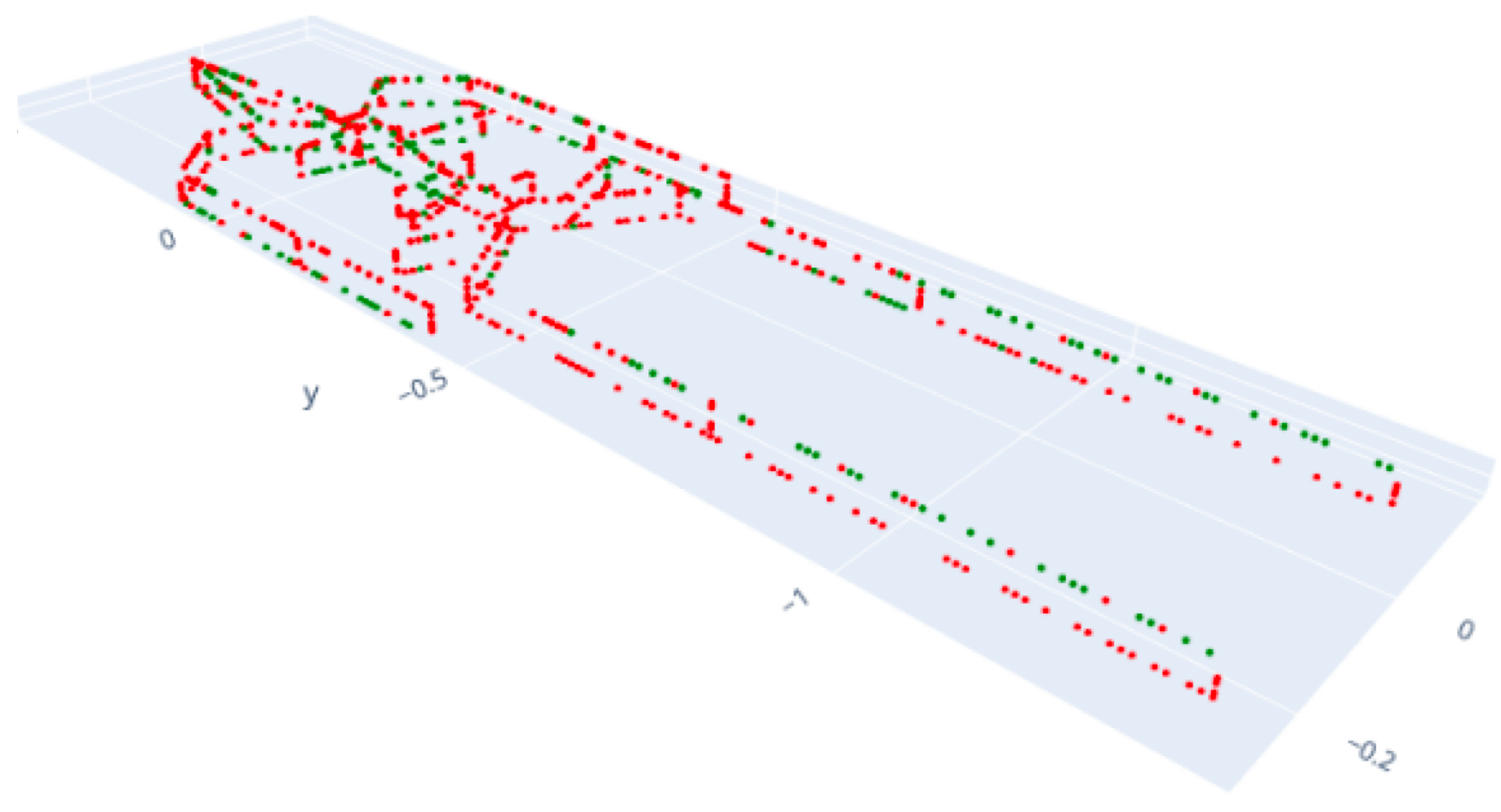}}
\subfigure[SRS]{
\includegraphics[width=0.238\linewidth]{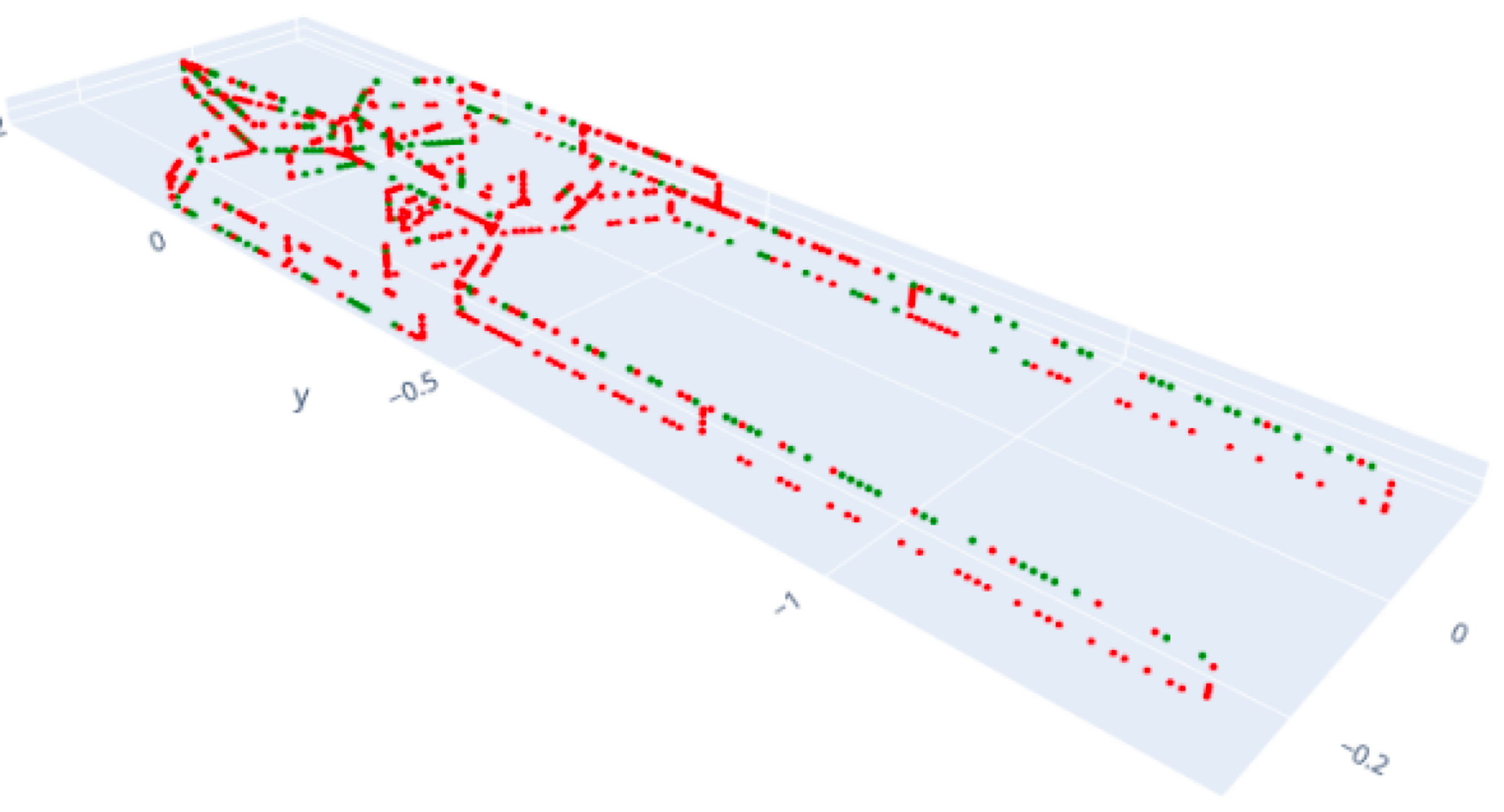}}
\vspace{-1mm}
\caption{Example regions accuracy distributions resulting from different sampling strategies}
\label{fig:sampling_visual}
\vspace{-3mm}
\end{figure*}

Table~\ref{tab:summary_results} overviews the average region and point accuracy achieved by utilizing different strategies. 
The considered strategies do not fluctuate significantly from the metrics averaged over all data target locations. 
The results show that \ac{RGS} converges fastest toward the overall performance. 
Specifically, it yields only a $\SI{0.2}{\percent}$ difference in region accuracy and $\SI{0.3}{\centi\meter}$ difference in point accuracy, compared to the baseline utilizing all sample locations. 
The average accuracy of region type 2 in the regular grid sample is $\SI{4}{\percent}$ lower than the overall accuracy. 
Nonetheless, the technique appears to produce consistent estimates of both region accuracy and mean error overall and for type 0. 
Although this strategy produces more accurate results than the other two, \ac{SRS} seems to generate more precise results for type 2 and \ac{SSRS} for type 1.
Example samples of target event locations obtained by utilizing different sampling strategies are depicted in Figure~\ref{fig:sampling_visual}.

Assuming that the goal is to assess the overall performance of a flow-guided localization solution, the simulator will employ \ac{RGS}. 
To guarantee that the obtained metrics are representative of the average performance in the cardiovascular system, at least \num{684} target locations, or $\SI{50}{\percent}$ of the overall dataset, should be utilized as a sample size. 
If the user is interested in specific region types, the strategy is based on the results in Table~\ref{tab:summary_results}.
Specifically, for region types 0, 1, and 2, the most adequate strategies would be \ac{RGS}, \ac{SSRS}, and \ac{SRS}, respectively.
This is despite the fact that for region type 1 \ac{RGS} outperforms \ac{SSRS} in terms of the point accuracy metric, however by a small margin. 
Hence, for the assessment targeting both metrics in region type 1 \ac{SSRS} is a marginally better sampling strategy than \ac{RGS}.

\begin{table}[!t]
\caption{Summary of performance convergence of selected sampling strategies}
\label{tab:summary_results}
\centering
\begin{tabular}{l c c}
\cline{1-3}
\textbf{Method} & \textbf{Region Accuracy [\%]} & \textbf{Mean Error [m]} \\
\cline{1-3}
Dense  & \hfil 28.6 & \hfil 0.210 \\
RGS    & \hfil \textbf{28.4} & \hfil \textbf{0.213} \\
SSRS   & \hfil 27.1 & \hfil 0.219 \\
SRS    & \hfil 26.9 & \hfil 0.205 \\
 \cline{1-3}
    & &  \\
\end{tabular}
\begin{tabular}{l c c c c c c}
 \cline{1-7}
\textbf{Method} & \multicolumn{2}{c}{\textbf{Region type 0}} & \multicolumn{2}{c}{\textbf{Region type 1}} & \multicolumn{2}{c}{\textbf{Region type 2}}  \\ \cline{1-7}
	  & \hfil Reg, & \hfil Error & \hfil Reg. & \hfil Error & \hfil Reg. & \hfil Error \\
	  \cline{2-7}
Dense & \hfil 31.8 & \hfil 0.235 & \hfil 30.6 & \hfil 0.179 & \hfil 13.4 & \hfil 0.228 \\
RGS	  & \hfil \textbf{31.5} & \hfil \textbf{0.236} & \hfil 33.1 & \hfil \textbf{0.175} & \hfil 9.4  & \hfil 0.249 \\
SSRS  & \hfil 29.3 & \hfil 0.237 & \hfil \textbf{29.4} & \hfil 0.186 & \hfil 14.0 & \hfil 0.256 \\ 
SRS	  & \hfil 29.5 & \hfil 0.231 & \hfil 29.2 & \hfil 0.171 & \hfil \textbf{13.8} & \hfil \textbf{0.221} \\ \cline{1-7}
\end{tabular}
\vspace{-1mm}
\end{table}

\subsection{Execution Times}

\begin{figure}[!t]
\centering
\includegraphics[width=0.93\linewidth]{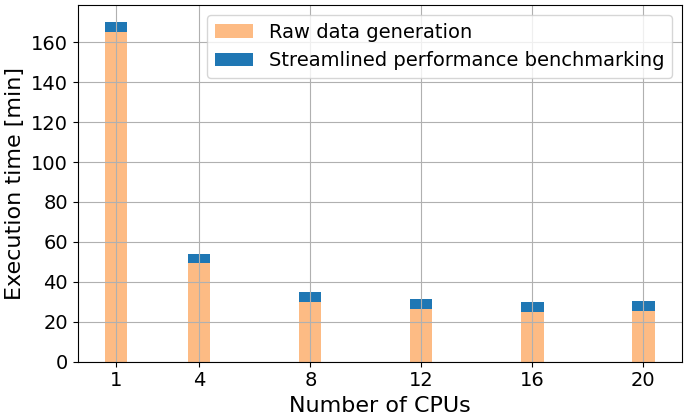}
\vspace{-1mm}
\caption{Example execution times of sequential and parallelized operation}
\label{fig:exec_times}
\vspace{-4mm}
\end{figure}

The proposed simulator provides support for the parallelization of benchmarking experiments. 
The reduction in the execution times of an example experiment due to parallelization is shown in Figure~\ref{fig:exec_times} for different \ac{CPU} configurations. 
In the example experiment, we have derived the raw data assuming \num{64} nanodevices deployed for $\SI{1200}{\second}$, while the performance metrics have been derived for \num{75} event locations in the cardiovascular system. 
As visible in the figure, parallelization can speed up the execution of experiments more than \num{6} times compared to the sequentially executed baseline. 
When running parallelized experiments, the experimenter should limit the CPU usage to maximum \num{12} units, as there are no significant benefits afterwards in terms of the reduction in execution times. 
Finally, one should observe that the majority of execution time is spent on the generation of raw data, and only a negligible constant duration of the streamlined performance benchmarking, as depicted in Figure~\ref{fig:exec_times}.
In other words, these results showcase that the design of our simulator allows for the generation of raw datasets representative of a variety of scenarios, on top of which fast and objective back-to-back performance benchmarking of different flow-guided localization approaches can be streamlined.

%% file: conclusion.tex
\section{Conclusion}
\label{sec:conclusion}

We argue that there is a need for objective evaluation of the performance of flow-guided nanoscale localization.
We further argue that such objectiveness can be achieved by utilizing the same evaluation environment, scenarios, and performance metrics.   
This is achieved by proposing a workflow for performance assessment of flow-guided localization and its implementation in the form of a simulator, providing the community with the first tool for objective evaluation of flow-guided localization\footnote{\url{https://bitbucket.org/filip_lemic/flow-guided-localization-in-ns3/}}.
Our results reveal relatively poor accuracy of the evaluated solution in the considered scenario.
Regardless of the poor accuracy, our results indicate that the proposed workflow and the implemented simulator can be utilized for capturing the performance of flow-guided localization approaches in a way that allows objective comparison with other approaches.
\textcolor{red}{This is based on the fact that we were able to interface two contemporary flow-guided localization approaches with the evaluation framework, obtain the estimations for both approaches utilizing the same raw data, and derive, present, and compare their performance results in a streamlined way.}

\textcolor{red}{One limitation of our work arises from the simplified representation of in-body nanodevice mobility compared to the complexity anticipated in real-world deployments of flow-guided localization systems. 
Specifically, the human bloodstream presents a much more intricate environment than the \acf{BVS} model utilized in this study. 
Future research should focus on enhancing the evaluation framework in terms of the complexity and precision of the pathways that nanodevices might navigate, as well as more accurately representing the complexities of blood flow. 
This includes better modeling of vortices and laminar flow, and accounting for blood vessel elasticity and bifurcations.
The modular implementation of the proposed simulation framework allows for potential integration of additional functionalities, as well as for substitution of existing ones.}

\textcolor{red}{Additionally, flow-guided localization is intended to be used in the bloodstreams of different individuals with varying biological characteristics. 
Future work will involve adapting the proposed framework to reflect individual variations in bloodstreams, similar to how anesthesia administration is tailored based on physiological indicators such as age, sex, height, and weight. 
For each patient, we will also consider temporal variations in the raw data stream for flow-guided localization due to factors such as physical activities, biological conditions (e.g., diseases), and environmental changes (e.g., temperature, humidity).
These in turn have an effect on the heart's pulsating rhythm, effectively changing the blood speeds.}

The poor accuracy of the considered approaches can be attributed to unreliable \acf{THz} communication between in-body nanodevices and on-body anchors and intermittent operation of the nanodevices due to energy-harvesting. 
Accuracy enhancements are envisioned along the lines of introducing additional anchors at strategic locations on the body (e.g., wrists) and developing a more suitable machine learning models that accounts for the fact that the raw data might be erroneous (e.g., compounding circulation times).
\textcolor{red}{For the development of suitable machine learning models, we consider \acfp{GNN} as the prime candidate due to their flexibility and resilient operation on graph structures~\cite{zhou2020graph}, which allows for accurately modelling the intricate dynamics of the nanodevices in the bloodstreams.
In that regard, it is worth pointing out to the recent work from Calvo \emph{et al.}~\cite{bartra2024graph}, in which a \ac{GNN}-based and multianchor-enabled flow-guided localization approach is proposed, featuring enhanced accuracy, reliability, and coverage compared to the benchmarks presented in this work.}